\documentstyle[prl,aps]{revtex}
\draft
\input epsf
\input{epsf.sty}
\begin{document}

\twocolumn[\hsize\textwidth\columnwidth\hsize\csname
@twocolumnfalse\endcsname

\title{
Possible Evidence for A Slush Vortex Phase in the Vicinity of Vortex Glass Transition in YBa$_{2}$Cu$_3$O$_{7-\delta}$ Thin Films
}

\author{
H. H. Wen$^{1}$, S. L. Li$^{1}$, G. H. Chen$^{2}$, and X. S. Ling$^{3}$}

\address{
$^1$National Laboratory for Superconductivity, Institute of Physics and Center for Condensed Matter Physics, Chinese Academy of Sciences, P.O. Box 603, Beijing 100080, China\\ 
$^2$Institute of Physics and Center for Condensed Matter Physics,
Chinese Academy of Sciences, P.O. Box 603, Beijing 100080, China\\
$^3$Department of Physics, Brown University, Providence, Rhode Island 02912, USA\\ 
}

\maketitle

\begin{abstract}
Current-voltage $I-V$ curves have been carefully measured for YBa$_2$Cu$_3$O$_{7-\delta}$ thin films by following different thermal or field annealing procedures. Although all data can be described quite well by the vortex-glass theory, it is found that the dissipation in a small region just above the vortex-glass transition is history dependent: the dissipation in the warming up or field-increasing process is larger than that in the cooling down or field-decreasing process. We attribute this history dependent dissipation to possible existence of a slush vortex phase.

\end{abstract}


]

The vortex matter continues to be an attractive subject in the study of high temperature superconductors (HTS). Due to the intrinsic properties ({\em e.g.}, small coherence length $\xi$, high operating temperature $T$, high anisotropy 
$\epsilon$, low charge density or high normal state resistivity $\rho_n$, {\em etc}. ) the HTSs have a rich and interesting vortex phase diagram which has received enormous effort in the past decade \cite{bla}. Much of the complexity of the vortex dynamics arises from the competing roles of vortex-vortex interactions, random pinning, and thermal fluctuations.   
In the absence of quenched disorder, the ideal vortex matter should have a crystalline ground state and can undergo a genuine solid-liquid phase transition upon heating.  In real type-II superconductors, however, the long-range translational order of the vortex lattice is destroyed by random impurities \cite{lo}.  Fisher, Fisher, and Huse \cite{ffh} conjectured that the low-temperature vortex phase in systems with random pinning could be a vortex glass (VG) characterized by the extremely non-linear $V(I)$ (or $E(j)$) dissipation \cite{fei}, or $E \sim exp[-(j_c/j)^\mu]$, and non-logarithmic magnetic relaxation \cite{thom}, consistent with many of the experimental findings \cite{koch,gam,chara,pet}.  In HTS films with strong random pinning centers, early experiments have shown a continuous transition with no hysteresis in transport properties \cite{koch}.  In this Letter, we report new results on similar HTS thin film samples using a more careful measurement procedure. Our data show clear evidence of a history dependent dissipation in the vicinity of the VG transition.  

Thin films measured for this work were prepared by magnetron sputtering deposition on (100) MgO substrates. 
The crystallinity of the films has been checked by the X-ray diffraction (XRD) pattern which shows only the (00l) peaks indicating a good epitaxy. The oxygen in the films has been depleted by annealing them in low pressure oxygen atmosphere, therefore the samples are slightly underdoped. The films with thickness of about 2000 {\AA} have been patterned into a bridge shape with width of 4 $\mu$m and length of 10 $\mu$m. After patterning, four silver pads were deposited and a post annealing at about 400 $^\circ C$ was made to reduce the contact resistance. (Joule heating is negligible for the current range of interest here.) The final bridge has a zero resistance temperature $T_{c0} = 86 K$ and transition width 
less than 1 K, the critical current density $j_c(77 K, H=0) = 2\times 10^6$ A/cm$^2$ using a voltage criterion of $E_c$ = 1 $\mu$V/mm. The transport measurement has been carried out with a Keithley 220 programable current source and a 182 nano-volt multimeter. An Oxford cryogenic system with magnetic field up to 8 Tesla was used to vary the temperature and magnetic field.  The temperature is controlled by adjusting the PID parameters of Oxford ITC4 tempearture controller with a temperature accuracy of about 0.02 K.

Fig. 1 shows the resistive transitions under several magnetic fields. It is found that the R(T) curves shift parallel to lower temperatures being similar to that in  
\begin{figure}[h]
    \vspace{10pt}	
    \centerline{\epsfxsize 8cm \epsfbox{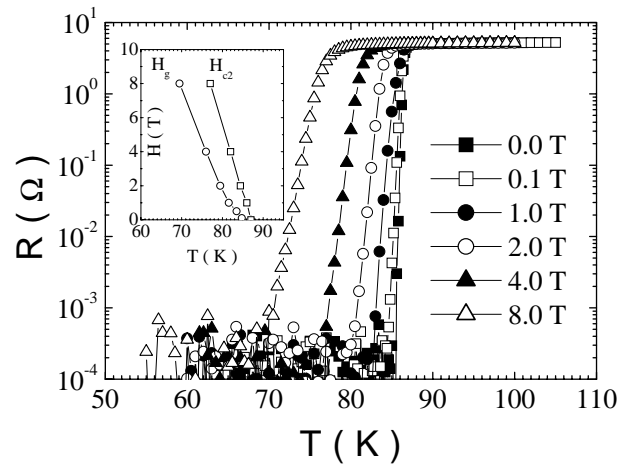}}
    \vspace{10pt}
\caption{Resistive transitions at magnetic fields ranging from 0 T to 8 T. The parallel shift of $R(T)$ is a property due to underdoping. Inset shows the upper critical field $H_{c2}(T)$ and the vortex glass transition line $H_g(T)$. }
\label{fig:Fig1}
\end{figure}
\noindent the underdoped region as observed by Suzuki et al.\cite{suzu} and Wen et al.\cite{wen1} This parallel shift of R(T) curve may be connected with the electronic properties in underdoped region, therefore we leave the discussion to a separate publication. The upper-critical field $H_{c2}(T)$ can be determined by extrapolating the flat line in the normal state and the steep transition portion to a cross point. As shown in the inset to Fig.1, the $H_{c2}(T)$ curve has a slope of about 0.8 T/K near $T_c(0)$. The irreversibility transition temperature $T_{irr}$ was determined by taking the criterion of $R =10^{-4} \Omega$. It is found that $T_{irr}(H)$ is close to the VG transition temperatures $T_g$ determined by extrapolating the curve of $dT/dlnR$ vs. $T$ to zero ( $dT/dlnR = 0$ ). Between $H_{c2}(T)$ and $H_{g}(T)$ it is supposed to be a 3D vortex liquid state separated from the VG state by the $H_{g}(T)$ line.

$I-V$ curves have been carefully measured at magnetic fields of 0.5 T, 1 T, 2 T, 4 T and 8 T.  For measuring these curves, four different annealing procedures are used: warming up ($T_{up}$) and cooling down ($T_{down}$) with fixed fields; field ascending ($H_{up}$) and field descending ($H_{down}$) at fixed temperatures. Special cautions were taken in these processes. In the $T_{up}$ ($T_{down}$) process, the sample is first cooled in a magnetic field to a temperature which is 10 K below (above) the target temperature,then it is heated up (cooled down) by stablizing the  temperatures in steps of 1 K up (down) to 1 K below (above) the target value. The next step is to approach the target temperature with an extremely slow speed by stablizing $T$ in steps of 0.1 K. In this process, over-heating or over-cooling should be avoided in preparing a reliable vortex state which ``records'' the history. In the $H_{up}$ ($H_{down}$) 
\begin{figure}[h]
	\vspace{10pt}
    \centerline{\epsfxsize 8cm \epsfbox{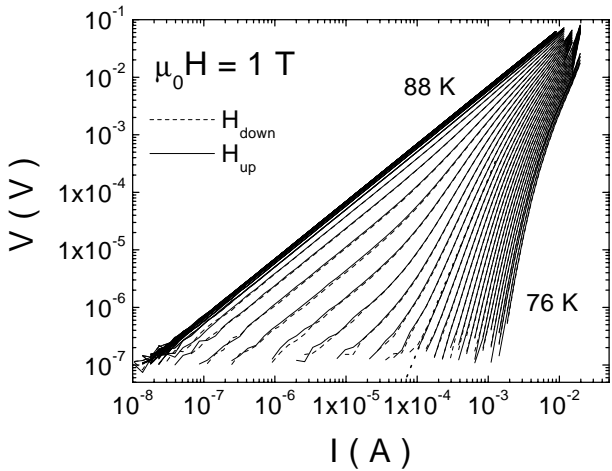}}
    \vspace{10pt}
\caption{$I-V$ curves measured in the so-called $H_{up}$ and $H_{down}$ processes (see text) from 76 K to 88 K with increments of 0.3 K. It is found that, in the temperature region just above $T_g$ ( marked here by the thick dotted line ) the dissipation in the $H_{up}$ process is always higher than that in the $H_{down}$ process showing a history dependent effect. It disappears in the high current (voltage) region for all temperatures.}
\label{fig:Fig2}
\end{figure}
\noindent process, the temperature is first fixed, the field is swept up (down) to the target field with a sweep rate of about 100 G/s.  The same cautions should be taken in order to prevent the field over-shooting.

Fig.2 presents the $I-V$ curves measured in $H_{up}$ (solid line) and $H_{down}$ (dashed line) processes. It is found that at temperatures below and well above $T_g$ marked here by the thick dotted line, the two sets of data coincide very well. In the small region just above $T_g$, it is surprising to see that, the $I-V$ curves measured in the $H_{up}$ process lies always above that measured in $H_{down}$ process indicating a history dependent dissipaption. When the current is increased into the region with a common negative curvature of $lnV$ vs. $lnj$, this effect disappears gradually. For a closer inspection to this history dependent dissipation, an enlarged view in the voltage region between $5\times10^{-6}$ V and $1\times10^{-4}$ V is shown in Fig.3.  This effect has been observed on the same sample with different runs of measurement by either varying the magnetic field at a fixed temperature or varying temperatueres at a fixed magnetic field. It is also repeatable on other YBa$_{2}$Cu$_3$O$_{7-\delta}$ thin films. Therefore it is a real effect. 

We find that, in spite of the striking history effects, the $I-V$ curves can nevertheless be shown to satisfy the scaling relations predicted by the VG theory.  According to Fisher {\em et al.}[3], at a second-order phase transition the $I-V$ curves can be scaled according to

\begin{equation}
\frac{V}{I{ \mid T-T_g \mid }^ {\nu (z+2-d)}} = f (\frac{I}{{T\mid T-T_g \mid}^{\nu(d-1)}})
\end{equation}
where $\nu$ and $z$ are critical exponents characterizing the VG transition, d is the dimensionality and f(x) is a universal function for all temperatures at a given magnetic field. In Fig.4 we present the VG scaling according to eq.(1) for the 
$I-V$ curves measured in the $H_{up}$ process at 1 T. Since the critical region for a VG scaling is  
\begin{figure}[h]
	\vspace{10pt}
    \centerline{\epsfxsize 8cm \epsfbox{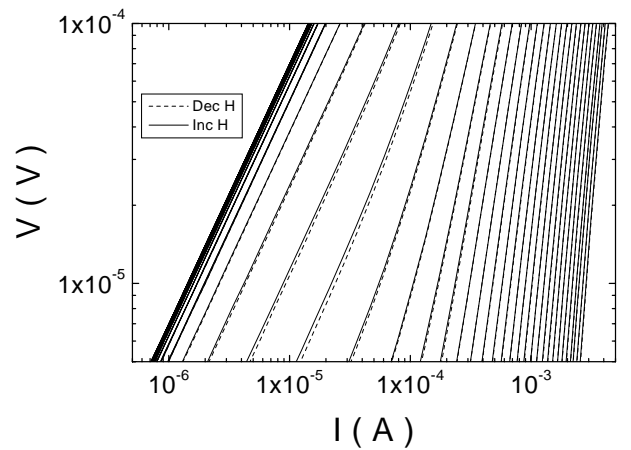}}
    \vspace{10pt}
\caption{An enlarged view of Fig.2 in the voltage region between $5\times10^{-6}$ V and $1\times10^{-4}$ V. The history dependent effect is more clear here.}
\label{fig:Fig3}
\end{figure}
\noindent narrow, therefore we use the data from 79 K to 85 K and voltage below $5\times10^{-3}$ V for the scaling. As shown by Fig.4, the data can indeed be scaled onto two main branches. The glass exponents obtained here are $\nu = 1.5$ and $z = 5$ which are very close to the values found by Koch {\em et al}. \cite{koch} in the same type of films. Another check is the ohmic resistivity just above $T_g$ which, in VG model, reads $ \rho = \rho_0(T-T_g)^s$ with $s = \nu (z-d+2)$, where $d$ is the dimensionality taken to be 3 for our case.  Taking $\nu = 1.5$ and $z = 5$ obtained from the scaling it is found 
that $s = 6$. In the inset to Fig.4 we show the data of ohmic $R(T)$. The $s$ value derived from the $R(T)$ data is 6.1 which is almost identical to that obtained from the scaling.  The VG scalings for the field of 2 T, 4 T and 8 T give almost the same glass exponents showing the generality of our films.  (It is important to point out that the apparent criticality in the $I-V$ characteristics could be caused by the correlated disorder such as grain boundaries and twins.  In that case, the universality class should be that of a Bose glass \cite{nel}.  For the purpose of comparison, we followed the same procedures as that of Koch {\em et al}.\cite{koch})  

To further investigate the history dependent effect, we take the difference between the currents corresponding to the same voltage in the increasing field (temperature) and deceasing field (temperature) processes at fields of 1 T and 4 T. As shown in Fig.5 (a) to (d), a clear peak appears in a narrow region just above the VG transition temperature $T_g$ marked by vertical thick dashed lines for all measurements. The open symbols: squares, circles and up-triangles represent the data taking at voltages of $5\times10^{-6}$ V, $5\times10^{-5}$ V and $5\times10^{-4}$ V, respectively. Clearly, the peak height due to this history dependent 
\begin{figure}[h]
	\vspace{10pt}
    \centerline{\epsfxsize 8cm \epsfbox{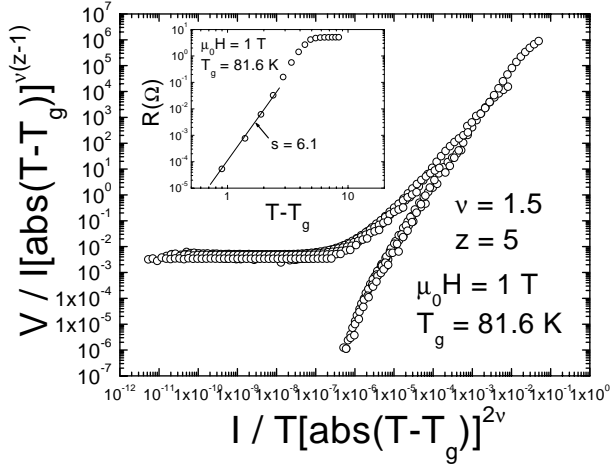}}
    \vspace{10pt}
\caption{Vortex glass scaling of the data shown in Fig.2. The same scaling has been taken to the data measured at other fields 1 T, 2 T, 4 T, and 8 T all yielding the similar values for $\nu$ and $z$. Inset shows the relation of resistance versus $T-T_g$ from which it is found that $s$ = 6.1 being very close to that determined from the scaling.}
\label{fig:Fig4}
\end{figure}
\noindent dissipation is supressed at a high voltage or current where the $logI$ vs. $logV$ shows a common negative curvature. This indicates that the history dependent effect appears in the so-called thermally-activated-flux-flow (TAFF) regime (ohmic). Another interesting finding is that the hysteretic behavior is stronger in the process of varying temperature than varying magnetic field.

The history dependent effect in transport properties has been observed in weak-pinning YBa$_{2}$Cu$_3$O$_{7-\delta}$ 
near the first-order phase transition \cite{saf}. In thin films with much stronger disorder, the transition is supposed to be second order \cite{ffh} and any hysteretic behavior is unexpected. Our data show that however even in thin films, the hysteretic behavior, although weak, can still be observed. This hysteresis may be understood by assuming the existence of a slush vortex phase just above $T_g$. The main argument is that the vortex matter with this hysteretic behavior is not in a single liquid state, rather it is in a liquid state containing some small vortex rystallites. \begin{figure}[h]
	\vspace{10pt}
    \centerline{\epsfxsize 8cm \epsfbox{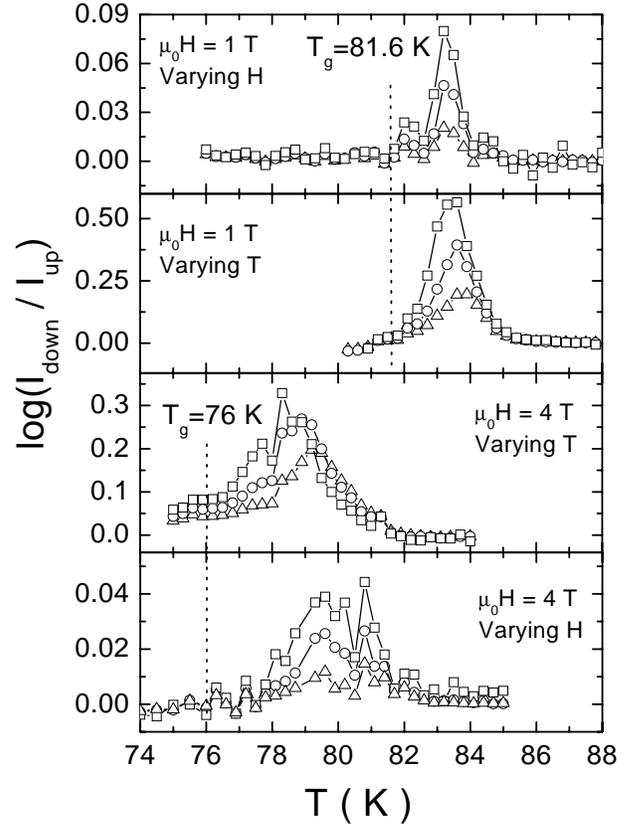}}
    \vspace{10pt}
\caption{The difference between the currents corresponding to the same voltage in the increasing field (temperature) and deceasing field (temperature) processes at fields of 1 T and 4 T. A clear peak appears in the small region just above $T_g$ marked here by the vertical thick dashed line. The open symbols: squares, circles and up-triangles represent the data taking at voltages of $5\times10^{-6}$ V, $5\times10^{-5}$ V and $5\times10^{-4}$ V, respectively. In the large voltage region, the peak is strongly suppressed.}
\label{fig:Fig5}
\end{figure}
\noindent When $T$ is below $T_g$, the vortex system is in the glassy state but with finite shear rigidity, therefore the small droplet or crystallites with translational order will play a role as building blocks for the whole elastic VG state. This picture is also consistent with the collective creep theory \cite{fei} that at a current smaller than the critical value, the optimal hopping object will be constructed by many Larkin domains. This will lead to a divergent high activation barrier in the small current limit. Above $T_g$, the strong thermal fluctuations will destroy the long-range phase coherence, leading to a second order melting. 
While just above the glass melting temperature, some small droplets or crystallites with short range order may survive. The melting of these small vortex cystallites may be first order and occur at a temperature slighly higher than $T_g$ causing the hysteresis effect. Since the vortices in the liquid state are softer than those in the small vortex crystals, they can easily adapt their configuration to find some strong pinning sites. Therefore when the temperature is lowered to the small region above $T_g$ the vortices are still in the plastic state, while when the temperature is increased to the same region from below $T_g$ some small vortex crystals remain unmelted. This will give rise to the slightly higher dissipation in the increasing temperature (field) process than in the decreasing process. Computer simulations by Dominguez et al.\cite{dom} have indicated that a large current can promote the melting of vortex lattice. This may explain in our present case why in the high current part the history dependent effect disappears. 

The term ``slush vortex'' was first introduced by Worthington {\em et al}.\cite{wor} in explaining the two step resistive transition in the defect-enhanced YBa$_2$Cu$_3$O$_{7-\delta}$ single crystals.  Recently Nishizaki {\em et al}.\cite{nishi} have found also in underdoped YBa$_2$Cu$_3$O$_{7-\delta}$ single crystals that the vortex glass transition $T_g(H)$ can be well below the critical point of melting line $T_m(H)$ showing the possible existence of a vortex slush regime between $T_g(H)$ and $T_m(H)$. Above two cases are close to the clean limit with single crystals as the beginning samples. Our present work reveals however that this slush vortex phase may also exist in exetremely disordered system, such as thin films. It would be interesting to know whether the history dependent dissipation found in our present paper occurs also in other systems, such as Bi$_2$Sr$_2$CaCu$_2$O$_8$ or Tl$_2$Ba$_2$CaCu$_2$O$_8$ since it has been shown \cite{wen2} that for such highly anisotropic systems, when the magnetic field is high enough, the vortex system melts at zero K with a non-zero linear resistivity at any finite temperatures.

In conclusion, by following different thermal and field annealing processes, a history dependent dissipation is observed in the vicinity of the vortex glass melting. This effect is attributed to the possible existence of a slush vortex phase due to the non-uniform distribution of the pinning sites.  

We acknowledge helpful discussions with Dr. Thierry Giamarchi and Dr. Xiao Hu.  This work was supported in part by the Chinese NSFC (projects: 19825111 and 10074078) and the Ministry of Science and Technology of China (NKBRSF-G1999064602). The work of XSL at Brown was supported by the NSF/DMR and the Alfred P. Sloan Foundation.


\end{document}